\documentclass[review]{elsarticle}

\usepackage{lineno,hyperref}
\usepackage{graphicx}
\usepackage{subcaption}
\usepackage{rotating} % Rotating table
\usepackage{amssymb}
\usepackage{amsmath}
\usepackage{amsmath}
\usepackage{amsfonts}
\usepackage{amssymb}
\usepackage{graphicx}
\usepackage{amsmath} 
\usepackage{algorithm}
\usepackage{adjustbox}
\usepackage{color}
\usepackage[noend]{algpseudocode}
\usepackage{multirow}
\usepackage{hhline}
\usepackage{hyperref}

\usepackage{graphicx}
\usepackage{caption}
\usepackage{subcaption}
\usepackage{epstopdf}
\usepackage{amssymb}
\usepackage{amsthm}
\usepackage{amsmath}
\usepackage{hyperref}
\usepackage{colortbl}
\usepackage{amssymb}
\usepackage{adjustbox}
\modulolinenumbers[5]

\usepackage{numcompress}

\begin{document}

\begin{frontmatter}

\title{Predicting Dynamic Modulus of Asphalt Mixture Using Data Obtained from Indirect Tension Mode of Testing}

% Group authors per affiliation:

\author[mymainaddress]{Parnian Ghasemi\corref{mycorrespondingauthor}}
\cortext[mycorrespondingauthor]{Parnian Ghasemi}
\ead[url]{pghasemi@iastate.edu}
\author[mysecondaryaddress]{Shibin Lin}
%\ead[url]{lin$\_$shibin@qq.com}
%\author[mymainaddress]{Jeramy C. Ashlock}
%\ead[url]{jashlock@iastate.edu}
\author[mythirdaddress]{Derrick K. Rollins}
%\ead[url]{drollins@iastate.edu}
\author[mymainaddress]{R. C. Williams}
%\ead[url]{rwilliam@iastate.edu}
%\author[mymainaddress]{Vernon R. Schaefer}
%\ead[url]{vern@iastate.edu}

%\author[mymainaddress]{Parnian Ghasemi\corref{mycorrespondingauthor}}
%\cortext[mycorrespondingauthor]{Parnian Ghasemi}
%\ead{pghasemi@iastate.edu}

\address[mymainaddress]{Department of Civil Engineering, Iowa State University, Ames, IA, USA}
\address[mysecondaryaddress]{Nondestructive Evaluation FHWA, Mclean, VA, USA}
\address[mythirdaddress]{Department of Chemical and Biological Engineering, Iowa State University, Ames, IA, USA}
\begin{abstract}
Understanding stress-strain behavior of asphalt pavement under repetitive traffic loading is of critical importance to predict pavement performance and service life. For viscoelastic materials, the stress-strain relationship can be represented by the dynamic modulus. The dynamic modulus test in indirect tension mode can be used to measure the modulus of each specific layer of asphalt pavements using representative samples. 

Dynamic modulus is a function of material properties, loading, and environmental conditions. Developing predictive models for dynamic modulus is efficient and cost effective. This article focuses on developing an accurate Finite Element (FE) model using mixture elastic modulus and asphalt binder properties to predict dynamic modulus of asphalt mix in indirect tension mode. An Artificial Neural Network (ANN) is used to back-calculate the elastic modulus of asphalt mixtures. The developed FE model was verified against experimental results of field cores from nine different pavement sections from five districts in the State of Minnesota. It is demonstrated that the ANN modeling is a powerful tool to back-calculate the elastic modulus and FE model is capable of accurately predicting dynamic modulus.
\end{abstract}

%\linenumbers
\end{frontmatter}
\section{Introduction}
\label{SEC:intro}
In order to predict pavement performance and service life, stress-strain behavior of asphalt pavement under repetitive traffic loading should be studied. One of the fundamental material properties of viscoelastic materials is the dynamic modulus which relates stress to strain \cite{kim2004dynamic,ghasemi2016performance,lin2016nondestructive,lin2015assessment,ashlock2015assessment}. Dynamic modulus is accepted by most pavement engineers as a critical parameter for flexible pavement design. As a stiffness indicator parameter, dynamic modulus is used in the Mechanistic-Empirical Pavement Design procedure \cite{Witczak2004guide,esfandiarpour2017alternatives,rahman2018assessment}.

The dynamic modulus test can be performed in two different test configurations, the uniaxial and indirect tension mode configurations. Both of these tests can be used in performance evaluation of the existing pavements. The uniaxial test configuration has the benefit of being simpler and less time consuming, while the indirect tension mode of testing is performed on a specific layer of a pavement section when the performance evaluation of a specific layer is of importance. Due to the multi-layer structure of the pavements, uniaxial text configuration is unable to evaluate each layer performance separately. Another issue with the uniaxial test configuration is the geometry requirement of the specimen. The test requires 6 inch tall specimens. Obtaining field cores with this dimension may not be possible in some cases with less than 6 inch thick Hot Mix Asphalt (HMA). Although indirect tension mode of testing can overcome this barrier, it is time consuming due to the manual tuning procedure and also the higher likelihood of damaging the specimens especially at higher temperatures. 

Developing predictive models for dynamic modulus is always important and several predictive models are currently being used widely \cite{Witczak2002simple}. The basic concept of these models is predicting dynamic modulus value based on material components' properties including binder, aggregate, and mixture volumetric properties \cite{Witczak2002simple,christensen2003hirsch,al2006new,junaid2018effect,yang2018use}. FE analysis is one way to develop such predictive models.

Finite element analysis has been used in modeling asphalt mixture with both elastic and viscoelatic assumptions \cite{breakah2015stochastic,herzog2008intrusive,Duncan1968finite,lytton1993development,ranadive2016parameter}. Researchers have developed two-dimensional and three-dimensional FE programs to simulate asphalt mixture behavior \cite{Witczak2002simple,collop2003development}. Finite element analysis is used to compare the material response with elastic and viscoelastic assumptions. Results indicate that the viscoelatsic assumption provides a more accurate estimation of pavement response \cite{elseifi2006viscoelastic,akbulut2005finite}. 
Although it has been shown that the finite element method is a powerful tool to simulate the viscoelastic behavior of asphalt mixtures \cite{breakah2009stochastic,zhang2017prediction,hu2017effects}, there has been a limited effort to use the application of FE modeling to simulate and further the understanding of the dynamic modulus test in the indirect tension mode.

In order to perform FE analysis, component properties of the pavement is required. Modeling viscoelastic behavior of asphalt mixtures requires defining both viscous and elastic properties of the material \cite{breakah2009stochastic}. Elastic modulus can be used to describe the elastic behavior of the pavement as an input in FE analysis. 
In the absence of the test results this property can be back-calculated using pattern recognition techniques. Modern pattern recognition techniques can learn and recognize trends in data contributing to their current widespread use \cite{ghanizadeh2018quasi,leiva2017non}. These techniques learn the pattern from experimental data and design the computational models. One such approach, Artificial Neural Network (ANN), is an interconnected network of many simple processors \cite{flood1994neural,ghasemi2018modeling,leiva2017non,ghanizadeh2017quasi,Ghasemi2018}. Figure 1 presents the schematic architecture of a network with three layers. All networks consist of a set of processing units or neurons classified as input, hidden, and output neurons. Input neurons receive inputs from external sources and transfer it to the rest of the network. Hidden neurons receive input and transmit their computed output to the processing units within the network without any outside contact. Output neurons receive the input from the rest of the network and transform and send it to external receivers.  The network can be trained by adjusting the network's weights and biases. Once the network learns the inputs/output(s) relationship, it can be used for further predictions \cite{flood1994neural}.

\begin{figure}
\centering
\includegraphics[width=0.75\textwidth]{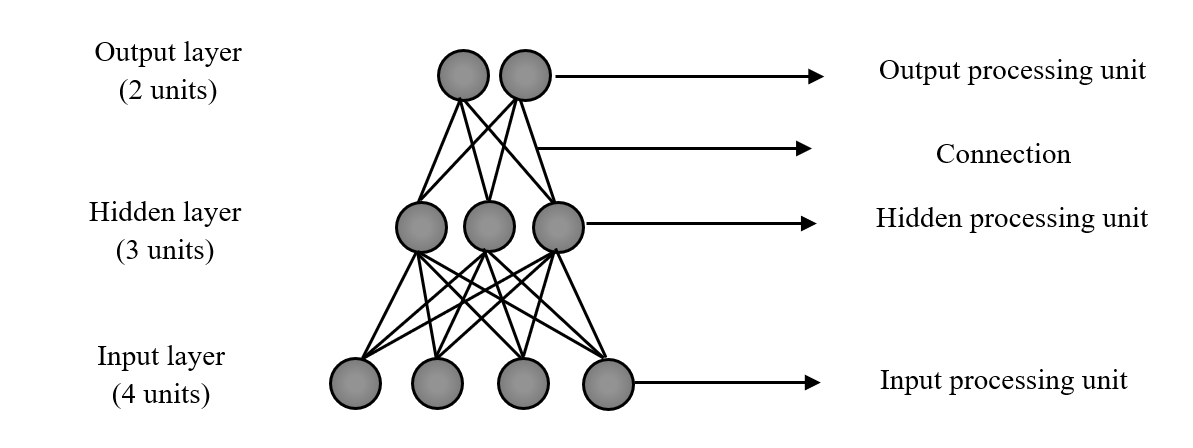}
\caption{Schematic architecture of a network with three layers}
\label{fig:1}
\end{figure}

This manuscript is organized as follows: the database creation procedure is presented in Section \ref{sec:2}. Section \ref{sec:3} covers the model development. Results and discussion  are presented in Section \ref{SEC:sec4}, followed by conclusions and recommendations in Section \ref{SEC:sec5}. 

\section{Database creation}
\label{sec:2}
Two separate databases are created for this research. The first one, database (a), is created using dynamic modulus values, complex shear modulus values, aggregate gradation, and mixture volumetric properties of twenty-seven field cores from nine different asphalt mixtures collected from five districts in the State of Minnesota. The summary information for each of these nine pavement sections is presented in Table \ref{TBL:1}.

\begin{figure}
\centering
\includegraphics[width=0.85\textwidth]{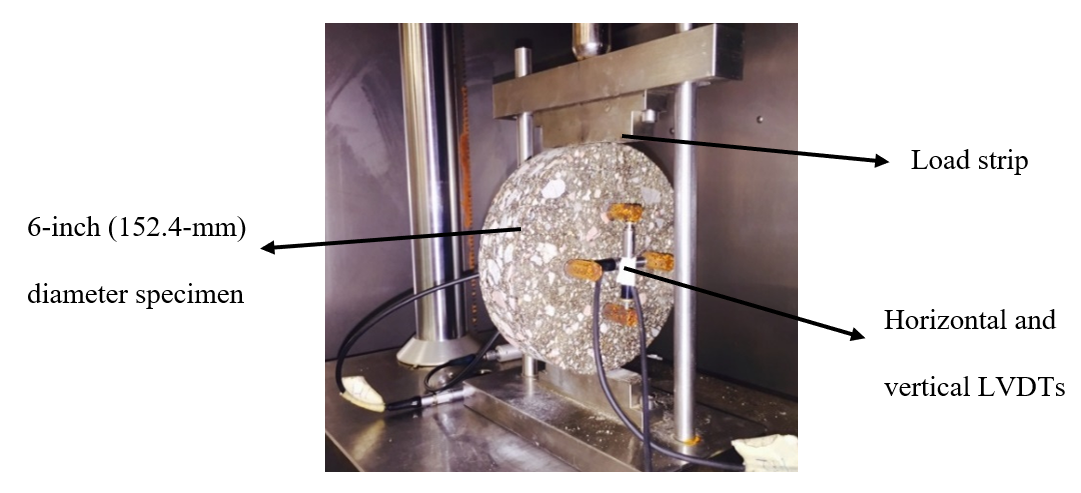}
\caption{IDT dynamic modulus test setup}
\label{fig:2}
\end{figure}

The field samples for dynamic modulus testing are 6 inches (152.4-mm) in diameter and about 1.5 inches (38.1-mm) in thickness. The dynamic modulus test in indirect tension mode is performed at three temperatures (0.4, 17.1, and 33.8 $^\circ C$) and nine loading frequencies (25, 20, 10, 5, 2, 1, 0.5, 0.2, 0.1 Hz)\cite{kim2004dynamic,Li.X2015practical}. Figure \ref{fig:2} illustrates the dynamic modulus test configuration in indirect tension mode of testing.

In order to obtain binder properties for each pavement group, binder extraction is done according to ASTM D7906(2014). The extracted binder then is recovered for testing following ASTM D2172(2011). A random number generator is used to select one sample from each pavement section for the binder extraction and recovery. The binder dynamic shear test is performed at the same test temperatures and loading frequencies as the mixture dynamic modulus test. It should be mentioned that unlike the modified Witczak model, in the present research a consistent definition of frequency is used \cite{sakhaeifar2015new}, that in order to predict the dynamic modulus value of an asphalt mix for example at 4 $^\circ C$ and 25 Hz, one should input in the model the complex shear modulus of asphalt binder, $|G^{*}|	$, at 4 $^\circ C$ and 25 Hz. 
Using the laboratory test results on 27 specimens, a database of 243 data points is created for subsequent use in the FE analysis.
The calculations are based on a linear viscoelastic solution for the IDT dynamic modulus test. All of the calculations are done based on the last five loading cycles. The applied sinusoidal loading can be expressed as shown in Eq.~\ref{EQ:eq1}

\begin{equation}
\label{EQ:eq1}
\centering
\begin{aligned}
P=P_0(cos \omega t+i sin (\omega t))
\end{aligned}
\end{equation}
where \(P_0\) is the amplitude of the sinusoidal load and \(\omega\) is the loading frequency.
Due to the load, a specimen will have a vertical and horizontal displacement which can be obtained from Eq.~\ref{EQ:eq2} and Eq.~\ref{EQ:eq3}, respectively 
\begin{equation}
\label{EQ:eq2}
\centering
\begin{aligned}
V(t)=V_0 sin(\omega t-\phi)
\end{aligned}
\end{equation}

\begin{equation}
\label{EQ:eq3}
\centering
\begin{aligned}
U(t)=U_0 sin(\omega t-\phi)
\end{aligned}
\end{equation}
where \(V_0\) and \(U_0\) are the constant amplitudes of vertical and horizontal displacement, respectively.
The dynamic modulus can be obtained from Eq.~\ref{EQ:eq4}

\begin{equation}
\label{EQ:eq4}
\centering
\begin{aligned}
|E^{*}(\omega)|=\frac{2|P_0|(\beta_1 \gamma_2-\beta_2 \gamma_1)}{(\pi a d)(\gamma_2 |\bar{V_0}|-\beta_2 |\bar{U_0}|)}
\end{aligned}
\end{equation}
where \(\beta_1\), \(\beta_2\), \(\gamma_1\) and \(\gamma_2\) are geometric coefficients which depend on the specimen size and gage length and are obtained from Eq.~\ref{EQ:eq5} to Eq.~\ref{EQ:eq8} \cite{hondros1959evaluation}.

\begin{equation}
\label{EQ:eq5}
\centering
\begin{aligned}
\beta_1= -\int_{-l}^{l} n(y)dy-\int_{-l}^{l} m(y)dy
\end{aligned}
\end{equation}

\begin{equation}
\label{EQ:eq6}
\centering
\begin{aligned}
\beta_2= \int_{-l}^{l} n(y)dy-\int_{-l}^{l} m(y)dy
\end{aligned}
\end{equation}

\begin{equation}
\label{EQ:eq7}
\centering
\begin{aligned}
\gamma_1= \int_{-l}^{l} f(x)dx-\int_{-l}^{l} g(x)dx
\end{aligned}
\end{equation}

\begin{equation}
\label{EQ:eq8}
\centering
\begin{aligned}
\gamma_2= \int_{-l}^{l} f(x)dx+\int_{-l}^{l} g(x)dx
\end{aligned}
\end{equation}

where,
\begin{equation}
\label{EQ:eq9}
\centering
\begin{aligned}
n(y)=tan^{-1} \left(\frac{1+\frac{y^2}{R^2}}{1-\frac{y^2}{R^2}} tan\alpha\right)
\end{aligned}
\end{equation}

\begin{equation}
\label{EQ:eq10}
\centering
\begin{aligned}
m(y)=\frac{(1-\frac{y^2}{R^2}) sin2\alpha}{1-2 (\frac{y^2}{R^2})cos2\alpha+\frac{y^4}{R^4}}
\end{aligned}
\end{equation}

\begin{equation}
\label{EQ:eq11}
\centering
\begin{aligned}
f(x)=\frac{(1-\frac{x^2}{R^2}) sin2\alpha}{1-2 (\frac{x^2}{R^2})cos2\alpha+\frac{x^4}{R^4}}
\end{aligned}
\end{equation}

\begin{equation}
\label{EQ:eq12}
\centering
\begin{aligned}
g(x)=tan^{-1} \left(\frac{1-\frac{x^2}{R^2}}{1+\frac{x^2}{R^2}} tan\alpha\right)
\end{aligned}
\end{equation}
where \(y\) and \(x\) are  the vertical and horizontal distances from the specimen center respectively, \(R\) is the specimen radius, \(\alpha\) is the radial angle, and \(l\) is half of the gage length.
  \begin{table}[]
\centering
\caption{Pavement sections information (* M/O = Mill and Overlay; ** O/L = Overlay)}
\label{TBL:1}
\begin{adjustbox}{width=0.8\textwidth}
\begin{tabular*}{1.3\textwidth}{ccccc}
\hline
Group No. & Section & MnDOT District & Construction Year & Construction Type           \\ \hline
1         & TH 220  & 2              & 2012              & 3'' M/O*                    \\
2         & CSAH 10 & 1              & 2012              & 1.5'' O/L** on old AC       \\
3         & TH27    & 3              & 2010              & 3'' M/O                     \\
4         & TH 9    & 2              & 2011              & 3'' O/L on reclaimed AC     \\
5         & TH 28   & 4              & 2012              & 4.5'' M/O                   \\
6         & TH 6    & 2              & 2010              & 1.5'' M/O                   \\
7         & TH 10   & 4              & 2013              & 3.5'' M/O                   \\
8         & CSAH 30 & Metro          & 2012              & 6'' M/O                     \\
9         & TH 10   & 3              & 2005              & 4'' M/O (sealed cracks)     \\
10        & TH 10   & 3              & 2005              & 4'' M/O (cracks not sealed) \\ \hline
\end{tabular*}
\end{adjustbox}
\end{table}              

The second database, database (b), is created using field cores taken from 20 different pavement sections in the States of Iowa, Wisconsin, and Minnesota. This database is used to back-calculate the elastic modulus based on the material components' properties. Dynamic modulus testing was performed on these specimens at four different temperatures and five loading frequencies. The interconversion method between linear viscoelastic material properties \cite{schapery1999methods} is used and presented in Eq.~\ref{EQ:eq13}
\begin{equation}
\label{EQ:eq13}
\centering
E' = |E^{*}|cos (\phi)
\end{equation}
where \(E' \) is the storage modulus,  \(|E^{*}|\) is the dynamic modulus, and\(\ \phi\) is  the phase angle. 
The storage modulus is an indicator of elastic behavior of the material and is used as an estimation of the elastic modulus. 
Aggregate gradation, asphalt binder shear properties, and volumetric properties of the mixture are retrieved from historical documents and a total of 240 data points for elastic modulus, aggregate, asphalt binder, and mixture properties are developed. 
\section{Model Development}
\label{sec:3}
\subsection{Back-calculation of the asphalt mixture elastic modulus in database (b)}

An ANN is an interconnected collection of processing elements \cite{saltan2002artificial} and can be trained to approximate a complex, nonlinear function through repeated exposure to produce meaningful solutions to the problem \cite{ghasemi2018modeling,saltan2007hybrid,aslani2010novel,rahami2011hybrid}. The multi-layer structure of the network ensures non-linear mapping of inputs to outputs.

A useful application of neural networks is in back-calculation procedures where it can be trained to approximate the inverse function by repeatedly showing it forward problem solutions. Once a network has learned the pattern of inputs/output(s)
relationship, it can predict new conditions \cite{flood1994neural}.

\begin{figure}[]
\centering
\includegraphics[width=0.7\textwidth]{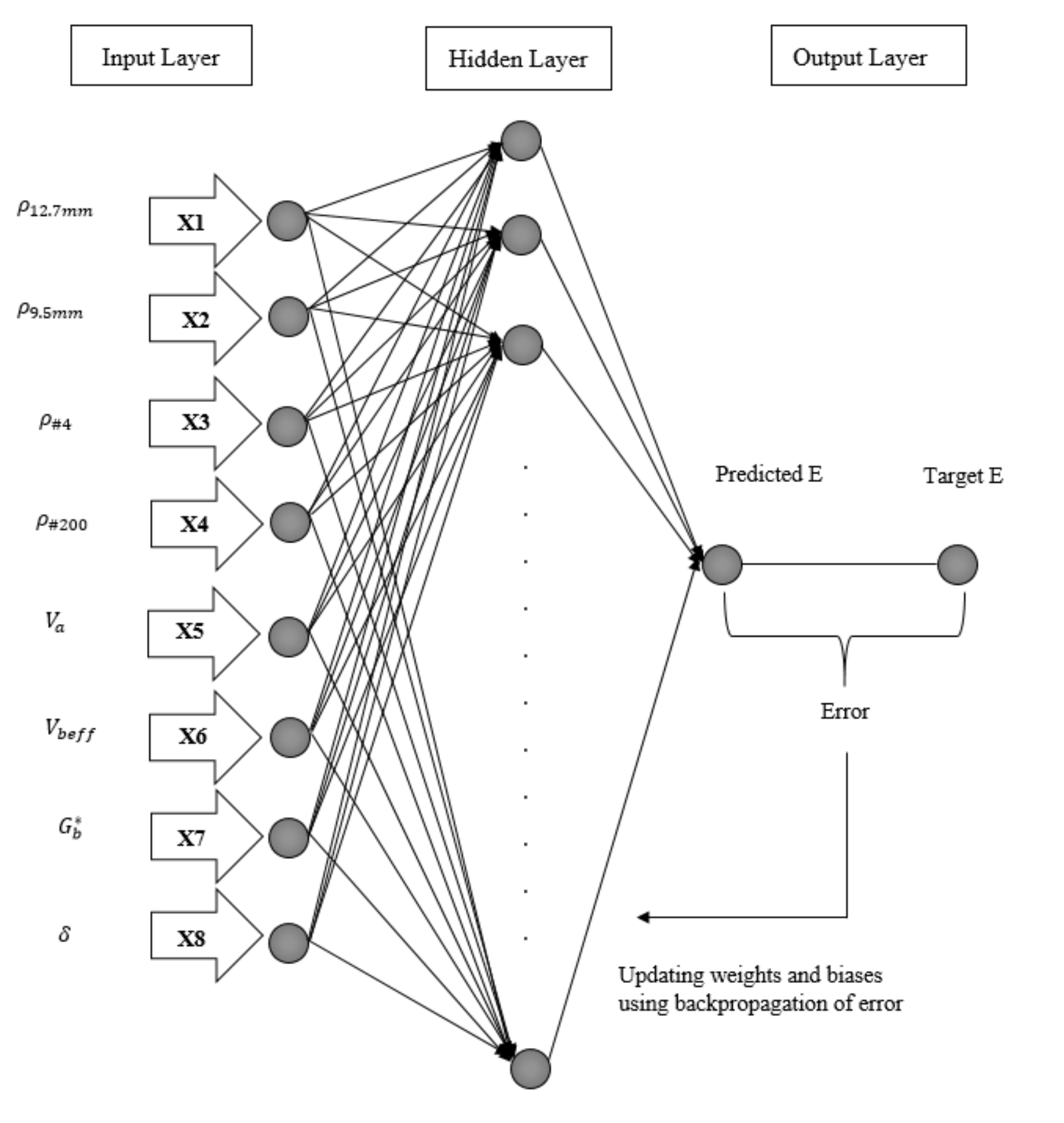}
\caption{Schematic architecture of a network with three layers}
\label{fig:3}
\end{figure}

Using database (b), a three-layer feed-forward error-back propagation network is trained. The network is composed of an input layer, an output layer and one hidden layer and developed using the MATLAB program \cite{matheworks2012matlab}. Figure~\ref{fig:3} presents the schematic architecture of the network having eight neurons (nodes) in the input layer, 10 neurons in the hidden layer and one neuron in the output layer. Selection of the number of hidden neurons is based on a trial-and-error optimization procedure balancing between the cost function and computational time. Selection of the input variables is based on the existing literature. There are several predictive models for dyanmic modulus that can predict the modulus value using material components' properties \cite{Witczak2004guide,al2006new,christensen2003hirsch,sakhaeifar2015new}. Researchers developed a network which is able to predict the value of dynamic modulus more accurate than modified Witczak model \cite{sakhaeifar2015new}. Their predictive models use the same input variables as modified Witczak model. The input variables are including aggregate gradation, binder shear properties, binder content, and air void. In the present study, same input variables are used with an additional step of converting dynamic modulus to the elastic modulus.

Database (b) is randomly divided into three categories. Seventy percent of the data points is presented to the network during the training procedure. The Levenberg-Marquardt method is used as the training algorithm \cite{more1978levenberg}. Fifteen percent of the data points is used for the validation of the fitting, and the rest of the data is used for testing. The training procedure starts with adjusting the initial values of the network's weights and biases to obtain a reasonable output and continue to modify the network by minimizing the mean squared residuals (MSE).
The input of each processing element \((\text{x}_{i})\) is multiplied by an adjustable connection weight (\(w_{\text{ij}})\). At each processing element, the weighted input signals are summed and a
threshold value \((B_{0})\) is added. This combined input is then passed through a non-linear transfer function \((f\left( . \right))\) to produce the output of the first layer \({(\nu}_{j})\), forming the input to the next layer. The network adjusts its weights on the presentation of a training dataset and uses a learning rule to find a set of weights that will produce the input/output mapping that results in the smallest possible error. This process is called ``learning'' or
``training''. In order to prevent overfitting, supervised learning procedure is used by means of having a validation subset. Once the training phase of the model has been successfully accomplished, the performance of the trained model must be validated using an independent validation set. After applying modifications and adjustments to the network's weights and biases the performance of the network will be examined by an independent testing set \cite{flood1994neural,cheng1994neural,juang1999cpt}. 

The output \(\nu_{j}\) from the $j^{th}$ hidden nodes is
given by
\begin{equation}
\label{EQ:eq14}
\centering
\begin{aligned}
\nu_{j} = f(x_{i},W_{ij}), \quad i=1,..., 8 \quad and \quad  j=1,..., 10
\end{aligned}
\end{equation}
and the single output \(\hat{y}\) is:
\begin{equation}
\label{EQ:eq15}
\centering
\begin{aligned}
\hat{y} = f_2 \left( f_{1} (\nu,W_{j}) \right).
\end{aligned}
\end{equation}
Then the expression of \(\hat{y}\) as a function of
\(\text{$x$\ }\) becomes a complicated nonlinear regression function with the \(j\) sets of weights, as parameters.
and for each j,
\begin{equation}
\label{EQ:eq16}
\centering
\begin{aligned}
\nu_{j} = f_{1} (B_{H_{j}}+W_{ij}x_{i}).
\end{aligned}
\end{equation}
So a general form of the feed forward neural network is described in
Eq.~\ref{EQ:eq17}:
\begin{equation}
\label{EQ:eq17}
\centering
\begin{aligned}
\hat{y} = f_{2} \left\lbrace B_{0} + \sum_{j=1}^{n} \left[ W_{j} \cdot f_{1} \left( B_{H_{j}} + \sum_{i=1}^{m} W_{ij} x_{i} \right) \right] \right\rbrace
\end{aligned}
\end{equation}
where \(B_{0}\) is bias at output layer (just one neuron at this layer),
\(W_{j}\) is weight of connection between neuron j of the hidden layer
and output layer neuron, $B_{H_{j}}$ is bias at neuron \emph{j} of
the hidden layer (for \(j = 1\ \text{to}\ 10)\), $W_{ij}$ is
weight of connection between input variable \emph{i} (for
\(i = 1\ \text{to}\ 8\)) and neuron \emph{j} of the hidden layer,
\(\text{x}_{i}\) is input parameter \emph{i},
\(f_{1}\left( t \right)\) is transfer function of the hidden layer, and
\(f_{2}\left( t \right)\) is transfer function of the output layer.

Both transfer functions \(f_{1}\left( t \right)\text{\ and}\ f_{2}(t)\)
used in this research are sigmoid functions as defined by Eq~\ref{EQ:eq19}:

\begin{equation}
\label{EQ:eq19}
\centering
\begin{aligned}
f_{k}(t) = \frac{1}{1+e^{-t}} \quad for \quad k=1,2.
\end{aligned}
\end{equation}

For the training set, the training procedure starts with adjusting the initial
values of the network's weights and biases to obtain a
reasonable output and continue to modify the network by minimizing the
value of the MSE. The iteration continues until the convergence
criterion is met as previously described. Once the network is trained and learns the pattern, it will be used to predict the elastic modulus values using database (a). The predicted elastic modulus will be used as an input in the finite element analysis.

\subsection{Finite element analysis}
\begin{figure}
\centering
\includegraphics[width=0.48\textwidth]{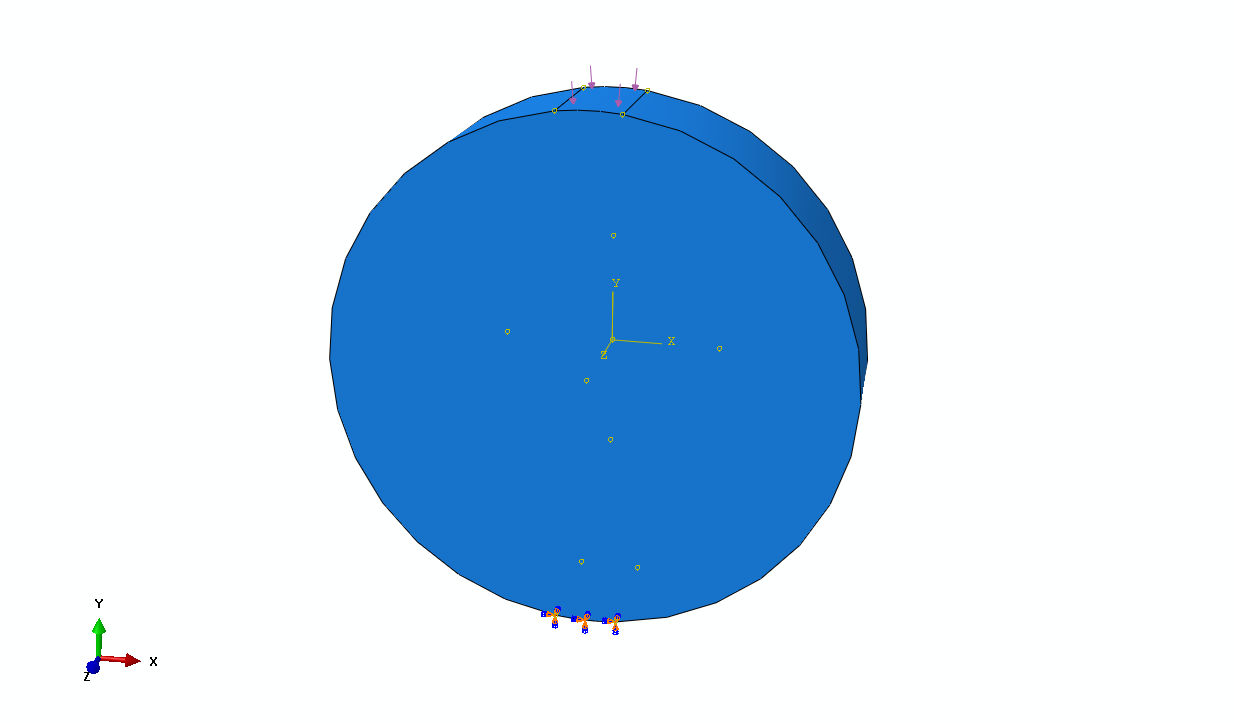}
\includegraphics[width=0.48\textwidth]{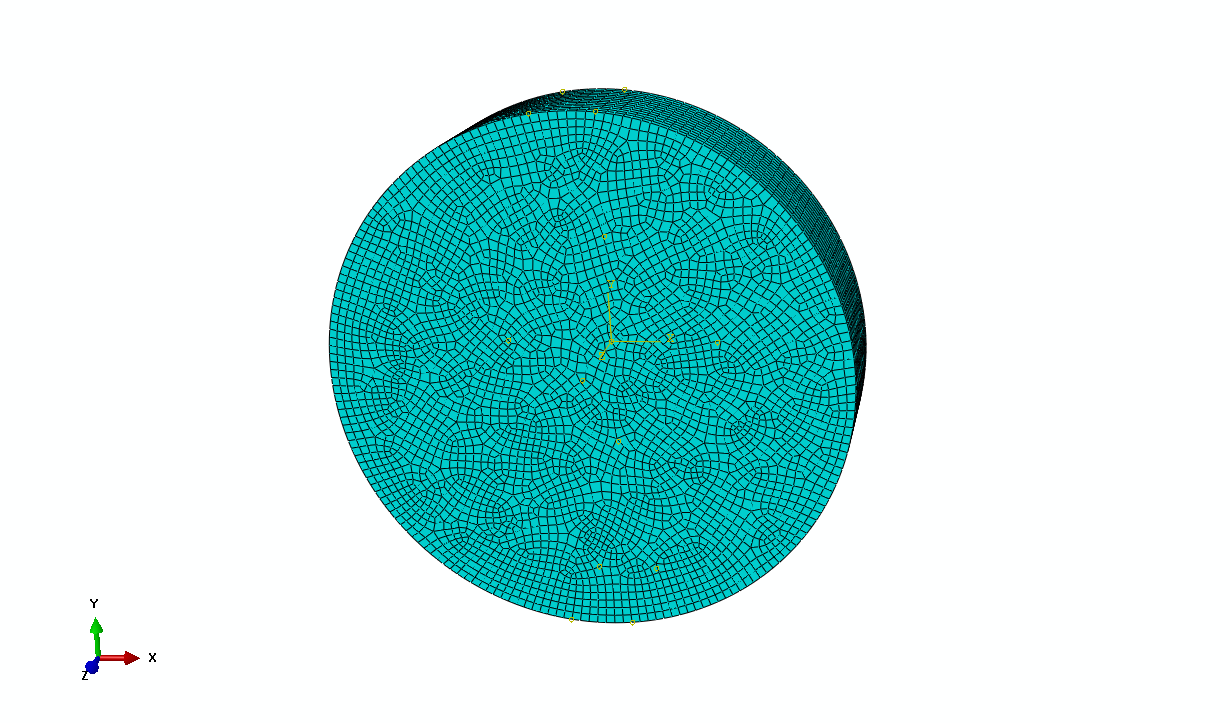}
\caption{Geometry, load, and boundary conditions (on the left), and mesh (on the right) used in FE analysis.}
\label{fig:4}
\end{figure}

Finite element analysis is performed using a multi-purpose FE software, ABAQUS \cite{abaqus20146}. ABAQUS has a module for viscoelastic materials which can be used for modeling asphalt mixture. A disk-shape geometry with the same dimensions as the laboratory HMA samples is created. The specimen is restricted at the bottom from movement and rotation in all directions and a sinusoidal uniform pressure is applied on top of the specimen via a loading strip. In order to determine the load magnitude, different load amplitudes are studied and adjusted so that the observed horizontal and vertical strains in the center area of the specimen remain between 60 and 80 micro-strain and below 100 micro-strain, respectively \cite{kim2004dynamic}. The mesh element type used for this model was from a 3D stress family, an eight-node linear brick, with reduced integration and hourglass control (C3D8R). Figure~\ref{fig:4} shows the model geometry, boundary conditions, and mesh.

There are several methods to define material properties in ABAQUS. Asphalt binder complex shear modulus test results are used in this paper. ABAQUS has a built-in function which transforms the shear modulus into a Prony series. The Prony series is an exponential expansion often used to describe the relaxation modulus of a viscoelastic material \cite{brinson2015characteristics}. The data is normalized as the ratio of the modulus at individual times to the long-term modulus.

The material temperature dependency should be defined in the model. Temperature dependency is defined using the shift factors obtained from Williams-Landel-Ferry (WLF) equation \cite{breakah2015stochastic,brinson2015characteristics,williams1955temperature} which is presented by Eq.~\ref{EQ:eq20}.

\begin{equation}
\label{EQ:eq20}
\centering
log(\alpha _{T})= \frac{C_1(T-T_s)}{C_2+T-T_s}
\end{equation}
where \(\alpha_T \) is the WLF shift factor, \(T\) is the temperature of each individual test, \(C_1\) and \(C_2\) are the constants and \(T_s\) is the reference temperature which is 17.1 $^\circ C$ in this research (Shear modulus values at 0.4 $^\circ C$ and 33.8 $^\circ C$ are shifted according to the reference temperature). For each individual temperature after determining \(C_1\) and \(C_2\) and shifting all of the complex modulus data according to the reference temperature, the value of these two constants at the reference temperature are used in the model as inputs.
Poisson's ratio is assumed to be 0.25. The selection of this value is based on material behavior at 17.1 $^\circ C$. A sensitivity analysis on the Poisson's ratio impact on the dynamic modulus value was conducted and the results demonstrated a very negligible effect exists.
As described earlier in the previous section, the elastic modulus of the material is predicted using the trained neural network. The predicted elastic modulus is used in the model as an input.

\section{Results and discussion}
\label{SEC:sec4}
Results of the laboratory tests, elastic modulus back-calculation, and finite element analysis are presented in this section. The capability and accuracy of the developed model in predicting dynamic modulus values is also evaluated. 

\subsection{Laboratory testing}
\begin{figure}
\centering
\includegraphics[width=0.87\textwidth]{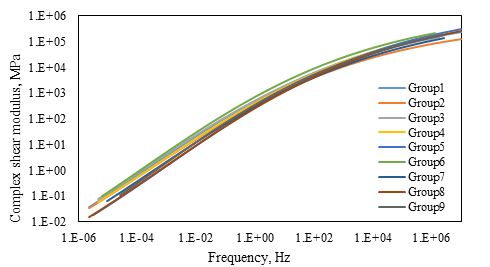}
\caption{Complex shear modulus master curves obtained from extracted and recovered asphalt binders of nine pavement sections}
\label{fig:5}
\end{figure}

The results of complex shear modulus test, dynamic modulus test, aggregate gradation, and mixture volumetric measurements are presented in this section. 
Complex shear modulus values are determined and are used to create master curves. The master curves for the nine groups are presented in Fig.~\ref{fig:5}.

To calculate dynamic modulus test results geometric coefficients, \(\beta_1\), \(\beta_2\), \(\gamma_1\), \(\gamma_2\), are calculated for specimen diameter of 152.4 mm (6'') and gage length of 65 mm (2.56'') and presented in Table~\ref{TBL:2}. The average dynamic modulus values for three specimens per group for the nine pavement groups are presented in Table~\ref{TBL:3}.

\begin{table}[]
\centering
\caption{Geometric coefficients}
\label{TBL:2}
\begin{adjustbox}{width=0.85\textwidth}
\begin{tabular*}{1.2\textwidth}{cccccc}
\hline
Specimen Diameter (mm) & Gauge Length (mm) & $\beta_1$     & $\beta_2$      & $\gamma_1$     & $\gamma_2$     \\ \hline
152.4                  & 65                & 0.0262 & -0.0078 & 0.0063 & 0.0206 \\ \hline
\end{tabular*}
\end{adjustbox}
\end{table}

\begin{table}[]
\centering
\caption{Average dynamic modulus values for nine pavement groups of database (a)}
\label{TBL:3}
\begin{adjustbox}{width=0.85\textwidth}
\begin{tabular}{ccccccccccc}
\hline
Group No. & Temp. & 25 Hz & 20 Hz & 10 Hz & 5 Hz  & 2 Hz  & 1 Hz  & 0.5 Hz & 0.2 Hz & 0.1 Hz \\ \hline
1         & 0.4   & 20425 & 20252 & 22008 & 18619 & 17501 & 16210 & 14836  & 13512  & 11995  \\
1         & 17.1  & 12573 & 12510 & 11315 & 9360  & 7537  & 6399  & 5081   & 3928   & 3104   \\
1         & 33.8  & 4798  & 4718  & 3462  & 2495  & 1680  & 1193  & 1000   & 656    & 502    \\
2         & 0.4   & 14822 & 17293 & 18841 & 17532 & 15858 & 14445 & 13132  & 11527  & 10191  \\
2         & 17.1  & 9768  & 9528  & 8905  & 7362  & 5751  & 4768  & 3934   & 2958   & 2386   \\
2         & 33.8  & 3157  & 2629  & 2077  & 1691  & 1203  & 982   & 779    & 631    & 593    \\
3         & 0.4   & 20128 & 19719 & 19727 & 18427 & 16927 & 15917 & 14710  & 13495  & 12285  \\
3         & 17.1  & 15137 & 15679 & 14442 & 12439 & 10730 & 9330  & 7999   & 6584   & 5652   \\
3         & 33.8  & 5769  & 5424  & 4419  & 3379  & 2353  & 1813  & 1394   & 1010   & 832    \\
4         & 0.4   & 21585 & 20264 & 19523 & 18049 & 16166 & 14767 & 13283  & 11639  & 10277  \\
4         & 17.1  & 13191 & 12485 & 11481 & 10044 & 7754  & 6457  & 5244   & 4008   & 3281   \\
4         & 33.8  & 5083  & 4877  & 3591  & 2718  & 1897  & 1443  & 1145   & 820    & 664    \\
5         & 0.4   & 22738 & 16279 & 16353 & 14882 & 13157 & 11970 & 10627  & 9074   & 7931   \\
5         & 17.1  & 9634  & 8889  & 7820  & 6369  & 4802  & 3875  & 3032   & 2244   & 1638   \\
5         & 33.8  & 3172  & 2882  & 2180  & 1621  & 1183  & 1005  & 950    & 750    & 607    \\
6         & 0.4   & 22324 & 23397 & 21829 & 20659 & 18753 & 17531 & 16515  & 14680  & 13266  \\
6         & 17.1  & 14264 & 13526 & 13312 & 11314 & 8720  & 7472  & 6167   & 4887   & 3921   \\
6         & 33.8  & 5512  & 5241  & 4146  & 3028  & 1995  & 1520  & 1138   & 766    & 557    \\
7         & 0.4   & 26413 & 22774 & 22624 & 21734 & 20130 & 18938 & 17544  & 15723  & 14373  \\
7         & 17.1  & 13950 & 13122 & 12836 & 10514 & 8153  & 6782  & 5470   & 4310   & 3459   \\
7         & 33.8  & 4486  & 4161  & 3377  & 2440  & 1658  & 1256  & 976    & 710    & 519    \\
8         & 0.4   & 24299 & 22946 & 22938 & 21377 & 20027 & 18331 & 16929  & 15092  & 13813  \\
8         & 17.1  & 12588 & 12151 & 10727 & 8796  & 7034  & 5822  & 4758   & 3653   & 2935   \\
8         & 33.8  & 4627  & 4006  & 3347  & 2695  & 1906  & 4886  & 1320   & 1171   & 1061   \\
9         & 0.4   & 20954 & 19559 & 20433 & 18470 & 16892 & 15498 & 14046  & 12499  & 11149  \\
9         & 17.1  & 11719 & 11584 & 10429 & 8377  & 6339  & 5194  & 4020   & 2965   & 2147   \\
9         & 33.8  & 4832  & 4195  & 3067  & 2171  & 1539  & 1198  & 919    & 614    & 468    \\ \hline
\end{tabular}
\end{adjustbox}
\end{table}

As described earlier, the results of dynamic modulus testing, complex shear modulus testing, aggregate gradation, and volumetric properties are used to develop database (a). Mixture volumetric properties are presented in Table~\ref{TBL:4}.

\begin{table}[]
\centering
\caption{Mixture volumetric propeeties for nine pavement groups of database (a)}
\label{TBL:4}
\begin{adjustbox}{width=0.8\textwidth}
\begin{tabular}{cccccccccc}
\hline
\begin{tabular}[c]{@{}c@{}}Group\\   No.\end{tabular} & 1     & 2     & 3     & 4     & 5     & 6     & 7     & 8     & 9     \\ \hline
\% RAP                                                & 23.8  & 23.3  & 37.2  & 26.2  & 23.8  & 36.4  & 23.3  & 11.4  & 45.3  \\
\% AC                                                 & 4.5   & 5.2   & 5.6   & 4.8   & 4.8   & 4.9   & 5.6   & 5.3   & 5.0   \\
\% Vbeff                                              & 4.2   & 4.1   & 4.1   & 3.9   & 3.5   & 4.3   & 4.2   & 4.0   & 4.6   \\
\%VMA                                                 & 13.5  & 13.5  & 13.6  & 13.1  & 12.5  & 13.9  & 13.7  & 13.4  & 14.4  \\
\% VFA                                                & 70.3  & 70.4  & 70.6  & 69.6  & 68.1  & 71.2  & 70.8  & 70.2  & 72.3  \\
Gmb                                                   & 2.315 & 2.315 & 2.315 & 2.315 & 2.315 & 2.315 & 2.315 & 2.315 & 2.315 \\
Gmm                                                   & 2.406 & 2.458 & 2.510 & 2.479 & 2.635 & 2.458 & 2.479 & 2.510 & 2.437 \\
\% VA                                                 & 4.010 & 3.996 & 3.998 & 3.982 & 3.988 & 4.003 & 4.000 & 3.993 & 3.989 \\ \hline
\end{tabular}
\end{adjustbox}
\end{table}

\subsection{Back-calculation of the asphalt mixture elastic modulus in database (b)}
The database (b) is created using elastic modulus (converted from dynamic modulus), asphalt binder properties, aggregate gradation, and mixture volumetric properties. The database is used to develop a neural network. A summary of the input variables used in the network and a descriptive measurement of them is presented in Table~\ref{TBL:5}.

\begin{table}[]
\centering
\caption{Network's input variables information}
\label{TBL:5}
\begin{adjustbox}{width=0.8\textwidth}
\begin{tabular}{ccccc}
\hline
\multirow{2}{*}{Variable} & \multicolumn{4}{c}{Values in the database} \\ \cline{2-5} 
                          & Maximum  & Minimum  & Average  & Std. Dev  \\ \hline
Complex Modulus (Mpa)     & 1065.6   & 0        & 45.4     & 117.2     \\
Phase angle (degree)      & 79.2     & 28.2     & 52.9     & 11.5      \\
Vbeff\%                   & 5.6      & 4.5      & 5.1      & 0.4       \\
Va\%                      & 4.0      & 4.0      & 4.0      & 0         \\
\% Passing 1/2''          & 96.4     & 87.2     & 93.9     & 2.6       \\
\% Passing 3/8''          & 87.3     & 73.7     & 81       & 4.1       \\
\% Passing \#4            & 63.8     & 48.2     & 54.1     & 5.3       \\
\% Passing \#200          & 6.2      & 3.1      & 3.8      & 0.9       \\ \hline
\end{tabular}
\end{adjustbox}
\end{table}

\begin{figure}
\centering
\includegraphics[width=0.9\textwidth]{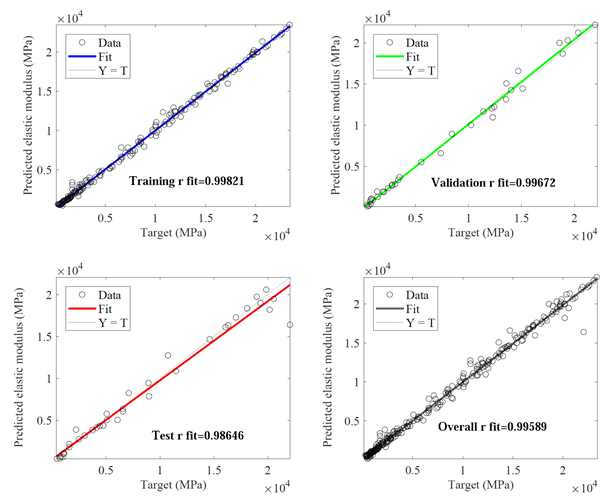}
\caption{Performance evaluation of the network. The top-left graph indicates performance of the network in fitting to the training subset with \(r_{fit}\)of 0.99. The top-right graph indicates performance of the network for validation subset with \(r_{fit}\)= 0.99. The bottom-left graph indicates the performance of the network for testing subset with \(r_{fit}\) of 0.98. The bottom-right graph indicates the overall performance of the network.}
\label{fig:6}
\end{figure}

As mentioned previously, the successfully trained three-layer ANN can be
presented as in Eq. \ref{EQ:eq17}. For ease of use and wider reproduction, the
connection weights and biases are presented using the following matrices:

\[W_{\text{ij}} = \begin{bmatrix}
 2.029  & 0.694  & -0.925 & 1.324  & 0.813  & 2.011  & -2.524 & -3.292 \\
-1.204 & 1.141  & 1.162  & 2.611  & 3.966  & 1.076  & -3.257 & 3.474  \\
2.135  & -1.270 & -0.277 & 0.179  & 0.683  & -1.184 & 1.174  & -3.481 \\
-0.549 & -4.837 & -2.383 & 0.794  & -2.491 & -1.766 & 1.930  & 5.408  \\
-0.777 & 1.006  & 2.255  & 0.436  & -1.330 & 1.137  & -2.658 & 4.933  \\
-0.757 & 1.685  & -1.889 & 1.674  & -3.114 & -2.335 & -6.051 & 3.408  \\
1.902  & -4.753 & 2.162  & -1.423 & 3.355  & 3.591  & -3.999 & -4.170 \\
-0.908 & 1.131  & 1.281  & 2.445  & 1.158  & 0.550  & 4.798  & -1.167 \\
2.171  & 0.007  & -0.669 & -1.475 & 1.944  & -0.571 & -4.329 & -2.316 \\
-0.684 & 1.626  & 0.741  & 0.582  & 0.041  & -1.464 & 1.508  & -0.188
\end{bmatrix}\]

\[W^{T}_{j} = \begin{bmatrix}
0.518  \\
-0.076 \\
0.614  \\
0.236  \\
0.082  \\
-0.244 \\
-0.280 \\
-1.012 \\
0.821  \\
2.000 
\end{bmatrix}\text{\ \ \ \ \ \ }{,\ B}_{\text{Hj}} = \begin{bmatrix}
 -4.851 \\
5.433 \\
-0.647 \\
3.163 \\
-1.864 \\
-6.336 \\
4.520 \\
-5.244 \\
4.0312 \\
-2.220
\end{bmatrix}\text{\ \ \ \ \ \ }{,\ B}_{0} = \left\lbrack 0.603
 \right\rbrack\]
Accuracy of the network is evaluated and the results are presented in this section. The performance evaluation is based on the $r_{fit}$ which is defined in Eq.~\ref{EQ:eq21}

\begin{equation}
\label{EQ:eq21}
\centering
\begin{aligned}
r_{fit} = \frac{n \sum_{i=1}^{n} E_{i} \hat{E}_i - (\sum_{i=1}^{n} E_i)(\sum_{i=1}^{n} \hat{E}_i )}{\sqrt{n\sum_{i=1}^{n} E^{2}_{i} - (\sum_{i=1}^{n}E_{i})^2}\sqrt{n\sum_{i=1}^{n} \hat{E}_i^2- (\sum_{i=1}^{n} \hat{E}_i)^2}}
\end{aligned}
\end{equation}
where \(r_{fit}\) is the correlation coefficient, \(E\) is the elastic modulus, and \(\hat{E}\) is the predicted elastic modulus.
Figure~\ref{fig:6} presents the performance of the network for training, validation and testing subsets in terms of \(r_{fit}\). The \(r_{fit}\) of 0.99 for testing, validation and testing subsets indicates that the ANN is capable of predicting the elastic modulus value of asphalt mixtures. The fact that the values of  \(r_{fit}\) for all of the subsets are close assures that the over-fitting is not likely to happen in the supervised training procedure.

\subsection{Finite element analysis}
\begin{figure}
\centering
\includegraphics[width=0.3\textwidth]{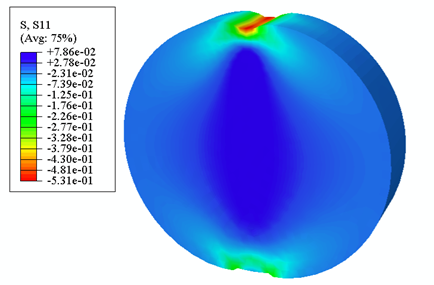}
\includegraphics[width=0.3\textwidth]{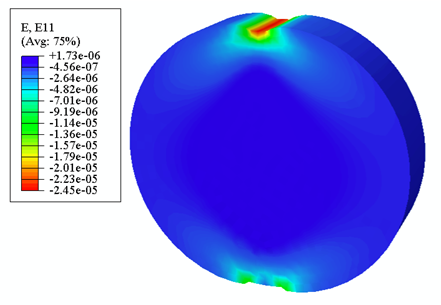}
\includegraphics[width=0.3\textwidth]{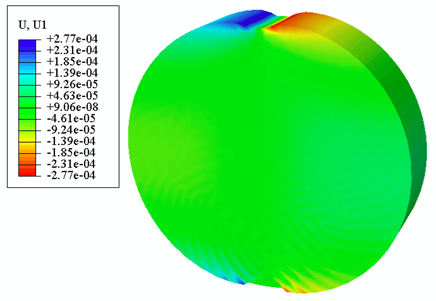}
\includegraphics[width=0.3\textwidth]{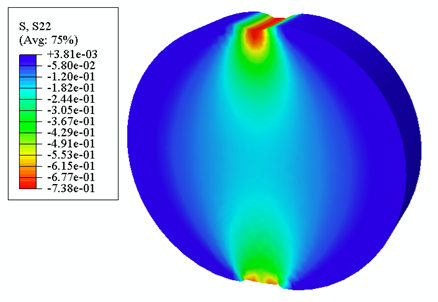}
\includegraphics[width=0.3\textwidth]{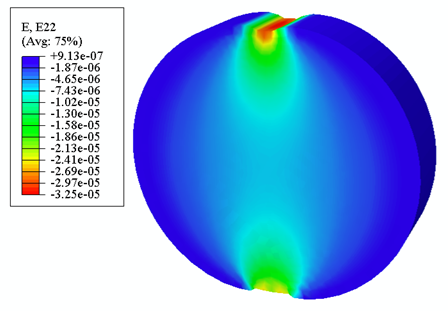}
\includegraphics[width=0.3\textwidth]{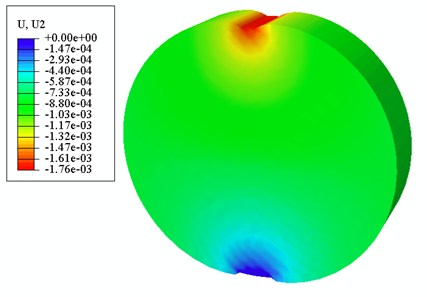}
\caption{Finite element analysis results. From left to right, the top ones indicate stress, strain and deformation in horizontal direction and the from left to right, bottom ones indicate stress, strain and deformation in vertical direction respectively.}
\label{fig:7}
\end{figure}

 Results of finite element analysis is presented in this section. The deformed shape of the HMA specimen due to sinusoidal loading with the loading frequency of 25 Hz is presented in Fig.~\ref{fig:7}.
 
\begin{figure}
\centering
\includegraphics[width=0.3\textwidth]{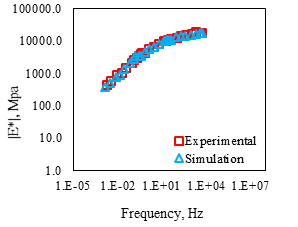}
\includegraphics[width=0.3\textwidth]{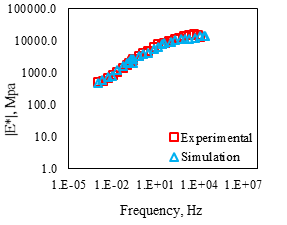}
\includegraphics[width=0.3\textwidth]{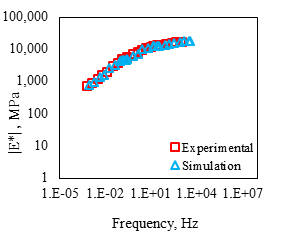}
\includegraphics[width=0.3\textwidth]{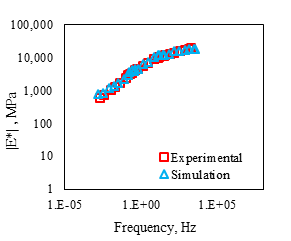}
\includegraphics[width=0.3\textwidth]{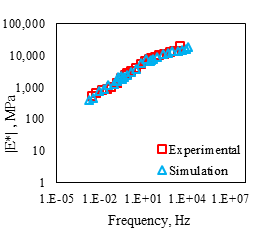}
\includegraphics[width=0.3\textwidth]{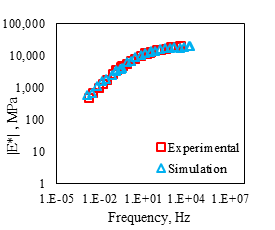}
\includegraphics[width=0.3\textwidth]{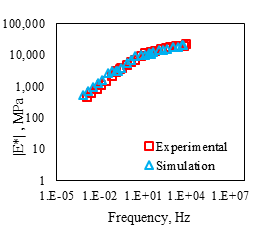}
\includegraphics[width=0.3\textwidth]{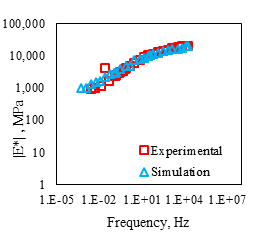}
\includegraphics[width=0.3\textwidth]{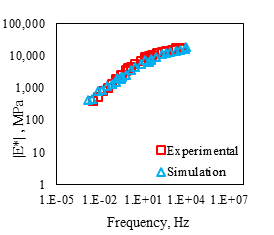}
\caption{Comparison of master curves created using experimental results and simulation results for 9 different pavement sections. Master curves are ordered from left to right so the very top-left one belongs to Group 1 and the very bottom one belongs to Group 9}
\label{fig:8}
\end{figure}

Vertical and horizontal stresses, \(S_{22}\) and \(S_{11}\), and deformations, \(U_2\) and \(U_1\), of the specimen under the applied load  are obtained for different loading frequencies. Complex modulus values for the nine different pavement groups are calculated for each frequency using Eq.~\ref{EQ:eq22}.

\begin{equation}
\label{EQ:eq22}
\centering
\begin{aligned}
E^{*}=\frac{S_{11}-\nu S_{22}}{E_{11}}
\end{aligned}
\end{equation}
where \(S_{11}\) is the horizontal stress along the x-axis, \(S_{22}\) is the vertical stress along the x-axis and \(E_{11}\) is the horizontal strain along the x-axis. The obtained modulus from Eq.~\ref{EQ:eq22} is the complex modulus of the material along the x-axis which is assumed to be not very different from the modulus along the y-axis. The dynamic modulus value is the amplitude of the complex modulus wave is presented in Eq.~\ref{EQ:eq23}

\begin{equation}
\label{EQ:eq23}
\centering
\begin{aligned}
E^{*}=|E^{*}| e^{(i \phi)}
\end{aligned}
\end{equation}
where \(i\) is the imaginary number and \(\phi\) is the phase angle.
The predicted values of dynamic modulus were calculated and compared with laboratory test results. Figure~\ref{fig:8} indicates a comparison between master curves obtained from experimental results and the ones obtained from the FE analysis. For all of the nine groups, the FE analysis was able to predict the dynamic modulus value.

\begin{figure}
\centering
\includegraphics[width=0.9\textwidth]{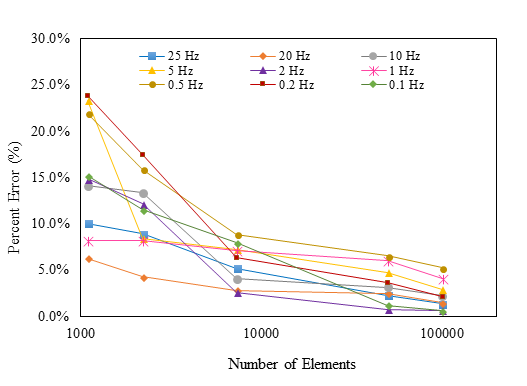}
\caption{Convergence study results for finite element analysis}
\label{fig:9}
\end{figure}

Element size is an important factor in determining the accuracy of solution in the FE method. To converge the solution to the correct value, the mesh should be properly discretized or a proper order element should be selected. In order to examine the ability of selecting a proper mesh by balancing accuracy of the solution as well as computational efficiency, a mesh convergence study is performed using five different mesh sizes including 10, 7.5, 5, 2.5 and 2 mm. The result of the convergence study for the first group of asphalt mixtures is presented. Figure~\ref{fig:9} represents a log-log scale of the relative error versus the number of elements for different loading frequencies. Based on the convergence study results, the relative error will converge to a specific value by increasing the number of elements or decreasing the mesh size.

The relative errors between simulation results and laboratory data for different loading frequencies and for different pavement sections were less than 20\%, which indicates that FE modeling is a capable tool to estimate dynamic modulus and performance evaluation of asphalt concrete. Larger amount of error was observed for groups 7, 8 and 9 which means that the assumption of homogeneity for asphalt mixture is not accurate although the predicted value is not far from the laboratory measurements. 
\subsection{Assessing the prediction performance: Receiver Operating Characteristics (ROC)}
\begin{figure}[]
\centering
\includegraphics[width=0.7\textwidth]{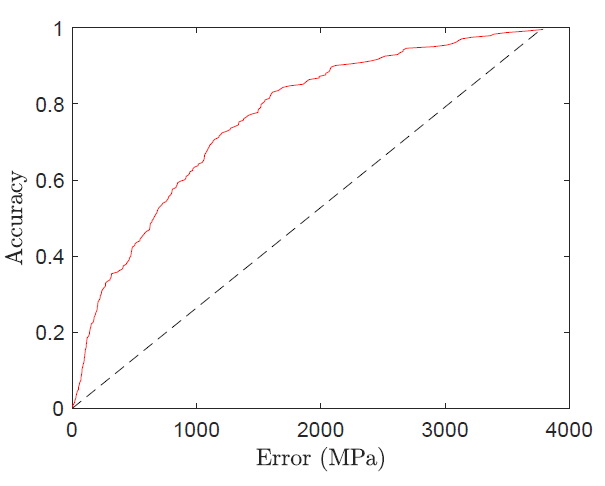}
\caption{Prediction accuracy with regard to a prespecified error tolerance}
\label{fig:10}
\end{figure}

A receiver operating characteristics (ROC) graph is a technique for visualizing, organizing and selecting classifiers based on their performance. The ROC graphs are used widely in medical decision making as well as machine learning and data mining research \cite{fawcett2006introduction}. The ROC curves plot the false positive rate on the x-axis and the true positive rate on the y-axis. A classifier performs well if the ROC curve climbs rapidly towards the upper left-hand corner. Random guessing on the other hand, will result in the diagonal line ($y=x$). The more the curve deviates from the $y=x$ behavior, the better the prediction is \cite{bi2003regression}.
The ROC curve can be used to visualize the performance of a predictive model by plotting the error tolerance on the x-axis and the accuracy of the prediction on the y-axis. Accuracy is defined as the percentage of points that are fit within the tolerance \cite{marti2013prediction}. In case of having zero tolerance, those points that the function fits exactly would be considered accurate. The obtained prediction accuracy is plotted for different error tolerances (margins) in Figure~\ref{fig:10}. The obtained curve is monotonically non-decreasing curve which climbs towards the upper left-hand corner (which is the desired situation) which shows that the predictive model performs well.

\subsection{Model validation}
\begin{figure}
\centering
\includegraphics[width=0.7\textwidth]{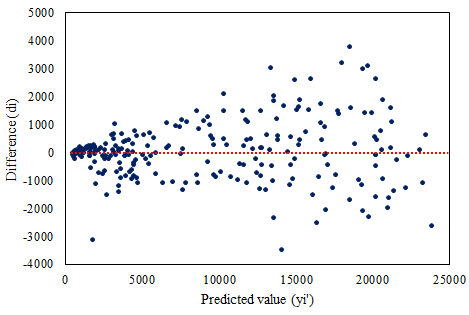}
\caption{Log of residual versus log of predicted values of dynamic modulus}
\label{fig:11}
\end{figure}

\begin{figure}
\centering
\includegraphics[width=0.7\textwidth]{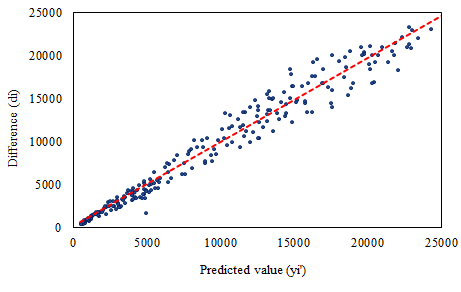}
\caption{Experimental results versus simulation results}
\label{fig:12}
\end{figure}

Considering each pair of the measured and predicted dynamic modulus values, their differences are obtained from Eq.~\ref{EQ:eq23}:
\begin{equation}
\label{EQ:eq23}
\centering
\begin{aligned}
d_{i} = y_{i}- {y^{'}_{i}}
\end{aligned}
\end{equation}
where \(y_{i}\) is the measured dynamic modulus value, \(y^{'}_{i}\) is the predicted values, and \(d_{i}\) is a random error term which assumed to be normally distributed with a mean of zero and
unknown variance \(\sigma^{2}\) for i = 1, . . ., n, where n is the number of input vectors. 
Many types of model inadequacies and violations of the underlying assumptions can be assessed. If the model is adequate, the residuals should contain no obvious pattern \cite{montgomery2017design}. The plot of the difference against the predicted values is presented in Figure.~\ref{fig:11}. Since there is no obvious pattern in this plot, the assumption of equal variances seems acceptable. 
Figure~\ref{fig:12} represents the linear relationship between experimental results and simulation results with the correlation coefficient of 0.98. 

\begin{figure}
\centering
\includegraphics[width=0.6\textwidth]{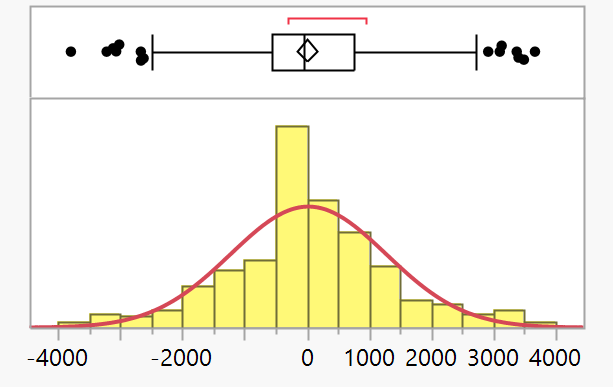}
\caption{A histogram of the differences,\(d_{i}\), with bell-shaped model overlaid}
\label{fig:13}
\end{figure}

\begin{figure}
\centering
\includegraphics[width=0.6\textwidth]{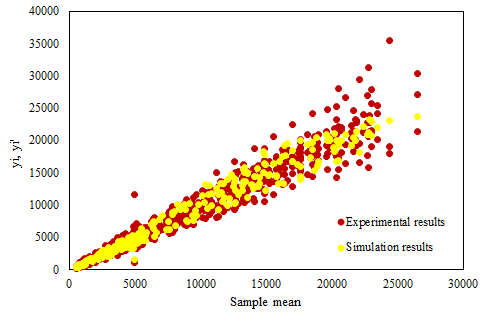}
\caption{Normal distribution of the difference between two methods}
\label{fig:14}
\end{figure}

A histogram with a bell-shaped model over the differences (\(d_{i}\)) is created and presented in Figure~\ref{fig:13}. The distribution appears to be more clustered around zero than normal plot. Experimental and simulated dynamic modulus values are ploted against their mean values in Figure.~\ref{fig:14}. According to this plot although spread increases as the order of the data increase, the spread of experiment is greater than the simulation spread or the simulation method is not adding any more variation to the existing variation caused by the experiment.
\section{Conclusions and recommendations}
\label{SEC:sec5}
The present study used the material components' properties to estimate the dynamic modulus value by means of finite element method (FEM).
Two separate databases were created for this research. The first one (database (a)) was created using twenty-seven field cores from nine different asphalt mixtures collected from five districts in the State of Minnesota.Database (a) contained dynamic modulus, binder shear properties, aggregate gradation and asphalt mix volumetric properties for 27 specimens.
The second database (database (b)) was created using field cores taken from 20 different pavement sections in the States of Iowa, Wisconsin and Minnesota.This database is used to back-calculate the elastic modulus based on material components' properties by means of an artificial neural network. The predicted elastic modulus along with binder shear properties were used as inputs in the finite element analysis. Simulation results were compared to the experimental data in terms of the dynamic modulus master curves. Based on this comparison, FE is a capable tool in predicting dynamic modulus of asphalt mixture. In the absence of field data. ANN is able to predict/back-calculate the elastic modulus of asphalt mixture. Larger database could be more liable to be used in such predictive modeling. Although running simulation instead of performing the laboratory testing is cost effective, developing predictive models by means of machine learning techniques could be even more efficient. A larger database with more variety (location, pavement mix design, etc.) should be used in future predictive modelings.

\bibliographystyle{elsarticle-num}

\bibliography{MyCollection}

\end{document}